\documentclass{article}

\usepackage[utf8]{inputenc}
\usepackage[margin=1in]{geometry}
\usepackage[titletoc,title]{appendix}

\usepackage{setspace}
\usepackage{bm}
\newcommand{\bb}{\bm}
\usepackage{latexsym,amssymb,color,fullpage}
\usepackage{mathtools}
\usepackage{amsthm}
\newcommand{\T}{\mathsf{T}}
\DeclareMathOperator{\Tr}{Tr}
\usepackage[ruled,vlined]{algorithm2e}
\newtheorem{theorem}{Theorem}

\usepackage{outlines}
\usepackage{enumitem}

\usepackage{amsfonts}
\usepackage{tabularx}
\usepackage[flushleft]{threeparttable}
\usepackage{booktabs}
\usepackage{subcaption}
\usepackage{setspace}
\usepackage{multirow}
\usepackage{comment}
\usepackage{rotating}

\usepackage{authblk}

\usepackage{amsmath,amsfonts,amssymb,mathtools}

\usepackage{graphicx,float}

\usepackage[ruled,vlined]{algorithm2e}
\usepackage{algorithmic}

\usepackage{minted}
\usepackage[style=authoryear, backend=biber, style=apa, natbib=true, bibstyle=apa, uniquename=false]{biblatex}
\usepackage[pdfencoding=auto, bookmarks=true, hidelinks]{hyperref}
\usepackage{hyperref, cleveref}
\hypersetup{
     colorlinks   = true,
     citecolor    = blue,
     linkcolor    = blue
}
\title{Reluctant Transfer Learning in Penalized Regressions for Individualized Treatment Rules under Effect Heterogeneity}
\author[1]{Eun Jeong Oh}
\author[2]{Min Qian}

\affil[1]{Northwell, New Hyde Park, NY 11042, United States}
\affil[2]{Department of Biostatistics, Mailman School of Public Health at Columbia University, New York, NY, 10032, United States}
\date{}

\begin{document}

\maketitle

\begin{abstract}

Estimating individualized treatment rules (ITRs) is fundamental to precision medicine, where the goal is to tailor treatment decisions to individual patient characteristics. While numerous methods have been developed for ITR estimation, there is limited research on model updating that accounts for shifted treatment-covariate relationships in the ITR setting. In practice, models trained on source data must be updated for new (target) datasets that exhibit shifts in treatment effects. To address this challenge, we propose a Reluctant Transfer Learning (RTL) framework that enables efficient model adaptation by selectively transferring essential model components (e.g., regression coefficients) from source to target data, without requiring access to individual-level source data. Leveraging the principle of reluctant modeling, the RTL approach incorporates model adjustments only when they improve performance on the target dataset, thereby controlling complexity and enhancing generalizability. Our method supports multi-armed treatment settings, performs variable selection for interpretability, and provides a regret bound for the difference in value of the optimal ITR and that of the estimated ITR. Through simulation studies and an application to a real data example from the Best Apnea Interventions for Research (BestAIR) trial, we demonstrate that RTL outperforms existing alternatives. The proposed framework offers an efficient, practically feasible approach to adaptive treatment decision-making under evolving treatment effect conditions.

\end{abstract}


\section{Introduction}\label{sec1}

Multi-site clinical trials often differ in the number of participants recruited at each site, which can complicate the estimation of individualized treatment rules (ITRs). This typically arise from variations in site capacity, patient availability, and recruitment rates. In the Best Apnea Interventions for Research (BestAIR) trial \citep{zhao2017effect, gleason2014challenges, yaggi2016reducing}, for example, one site enrolled only half the number of patients compared with another. Simply pooling data across sites may obscure meaningful site-specific treatment effects, while fitting models separately may yield unstable estimates for sites with limited sample sizes. These challenges motivate borrowing information from a larger, informative source site to improve model fitting on target sites, which can ultimately lead to more accurate estimation of ITRs for each individual within each site.
Numerous methodological frameworks have been proposed for the ITR estimation from randomized clinical trials or observational studies (e.g., \citet{murphy2003optimal, robins2004optimal, zhao2012estimating, zhang2012robust, qian2011performance}).
However, most of these approaches rely on the assumption that the observed data comes from a single population in which the relationships between covariates, treatments, and outcomes are estimated uniformly. In many real-world applications, it is often necessary to develop site- or study-specific models because treatment effects can vary across settings. Another common situation arises when models are developed using historical (source) data and later confronted with new (target) data as more patients are enrolled or studies are expanded.

While transfer learning has gained popularity and provided a promising framework for addressing these challenges, most existing approaches for the ITR estimation have primarily focused on covariate shift, where the distribution of covariates differs across populations while the conditional treatment effect is assumed to remain stable \citep{zhao2019robust, wu2023transfer, mo2021learning, chu2023targeted, sui2025robust}. They commonly adopt reweighting or robust optimization techniques to mitigate distributional changes, but do not account for effect heterogeneity. 
When such effect heterogeneity is present, relying solely on covariate alignment may yield biased or unstable estimates.


Recent studies have emphasized the importance of approaches that explicitly account for heterogeneity in treatment effects across data sources \citep{dahabreh2020toward, robertson2023regression, kunzel2019metalearners}. This consideration is particularly critical for estimating ITRs, where shifts in treatment-covariate interactions can lead to suboptimal decisions for patients if not properly addressed. 
Although not directly related to ITR, several studies have recognized that regression coefficients may vary across subgroups. For instance, \citet{gross2016data} investigated the identification of sub-populations in randomized trials for which an intervention is beneficial, while \citet{ollier2017regression} studied the identification of strata likely to share similar effects. More recently, \citet{li2022transfer} explored the multi-source high-dimensional linear regression problem to improve a model fit on target data through a two-step procedure. Although the study offers valuable theoretical insight, it provides limited practical guidance for interpretable or decision-oriented applications.


There remains a gap in transfer learning approaches that are both efficient and clinically intuitive, while directly addressing the problem of shifted treatment-covariate relationships in the ITR setting. {\color{black} To address this gap, we propose a framework that adapts a source model to new target data by allowing it to flexibly deviate from the source model, through a pseudo-outcome construction to recover target estimators and identify shifted treatment effects. By transferring only essential information, such as regression coefficients, our approach remains computationally efficient while preserving interpretability. Specifically,} we draw inspiration from the principle of reluctant modeling to transfer information and allow for ``shifted effects.'' Such shifted effects are incorporated only if they enhance predictive performance, inspired by reluctant modeling \citep{yu2019reluctant, tay2020reluctant}. Recently, \citet{maronge2023reluctant} applied this principle to ITR estimation by introducing a reluctant additive model that includes additional terms only if they improve prediction. However, their work was not designed to handle model adaptation across datasets and is limited to binary treatments.

In the present work, we propose the Reluctant Transfer Learning (RTL) method that integrates a previously trained model with newly collected target data by transferring only essential model components (e.g., regression coefficients) rather than individual-level data from the source samples. 
The RTL method directly addresses the need for model adaptation and is able to handle  multi-armed treatment settings. In addition, we incorporate penalization techniques for variable selection to enhance interpretability and performance. 
By combining the strengths of reluctant modeling and transfer learning, the RTL framework offers a statistically rigorous and clinically interpretable approach that is well suited for precision medicine applications in which models must adapt when necessary.

The paper is organized as follows. In Section \ref{sec2}, we introduce a general framework of obtaining optimal ITR with the proposed RTL method, and we also present a regret bound for the difference in value of the optimal ITR and that of the estimated ITR.
In Sections \ref{sec3}, we compare our proposed method with other existing alternatives through simulation studies. In Section \ref{sec4}, we apply our RTL method to a real data example from the BestAIR trial.  
Discussion and conclusions are presented in Section \ref{sec5}. The proof of theorem is included in Appendix.

\section{Reluctant Transfer Learning in \texorpdfstring{$\ell_1$}{l1}-Penalized Regressions} \label{sec2}

Denote the source data as $(Y_{s}, \bb O_{s}, \bb A_{s})$ from $n_s$ samples, where $\bb O_s$ is the vector of individual covariates, $\bb A_s$ denotes treatments, and $Y_s$ is the outcome of interest with large values desired. We assume that $\bb A_s$ involves discrete treatments. {\color{black} When more than two treatment options are available, $\bb A_s$ is encoded as a vector of dummy variables.}
Similarly, the target data $(Y_t, \bb O_t, \bb A_t)$ are observed from $n_t$ samples, following the same structure and definitions.
In the context of target samples, an individualized treatment rule (ITR), $\bb\pi_t$, is a mapping from the space of observations,  $\mathcal{O}_t$, to the space of treatments, $\mathcal{A}_t$. The {\it value} of the decision rule, denoted as $V(\bb\pi_t)$, is the expected outcome that would be obtained if the decision rule is implemented in the future. Our goal is to estimate an optimal ITR on the target samples, $\bb\pi_t^o$, that would maximize the expected outcome if implemented:
\begin{align*}
\bb\pi_t^o = \arg\max_{\bb\pi_t} V(\bb\pi_t).
\end{align*}

Throughout, $E$ denotes the population average and $E_n$ denotes the empirical expectation over the observed sample. For notational simplicity, we use $n$ in place of $n_t$ whenever appropriate; for example, $E_n$ applied to target samples represents the sample average across $n_t$ observations.
Let $\bb\Phi_{t} \in \mathbb{R}^p$ is a vector summary of $(\bb O_{t}, \bb A_{t})$. 
Define the $Q$-function $Q( \bb \Phi_{t}) = E(Y_{t} |  \bb \Phi_{t})$ such that $Q(\bb o_t, \bb a_t)$ measures the quality of assigning treatment $\bb A_t = \bb a_t$ to an individual with $\bb O_t = \bb o_t$ \citep{qian2011performance, murphy2005generalization}. The model on the target data $(Y_{t}, \bb O_{t}, \bb A_{t})$ is defined as
\begin{align}
Q (\bb\Phi_{t}) = \bb \Phi_{t}^\T \bb\beta_{t},
\label{model}
\end{align}
where we assume $\bb \beta_{t} = \bb \beta_{s} + \bb \theta$ such that $\bb \beta_{t}$ shares some similarity with $\bb \beta_{s}$ to some extent by the additive parameter $\bb \theta$. Herein, the parameter $\bb{\beta}_{s}$ represents a linear relationship between $\bb\Phi_{s}$ and the corresponding $Q$-function on the source data, and the parameter $\bb\theta$ represents a shift in the vector of coefficients between the model previously trained on the source data and the new model for the target samples. 

We consider the case in which there exists a well-trained pre-model based on the source data, primarily represented by $\hat{\bb{\beta}}_{s}$, which may be any root-$(n_s/p)$ consistent estimator. 
Our proposed Reluctant Transfer Learning (RTL) passes $\hat{\bb{\beta}}_{s}$ to construct the pseudo-outcome $\tilde Y_{t} = Y_{t} - \bb \Phi_{t}^\T \hat{\bb\beta}_{s}$ and then obtains $\hat{\bb\theta}$ by minimizing the following objective function
\begin{align}
n_t E_n(\tilde Y_{t} - \bb \Phi_{t}^\T \bb\theta)^2 + \lambda_n \sum_{j=1}^{p} w_j |\theta_j|,
\label{obj}
\end{align}
and it finally recovers the estimators on the target samples by $\bb{\hat\beta}_{t} = \bb{\hat\beta}_{s} + \bb{\hat\theta}$. Accordingly, the estimated ITR is the intervention that maximizes the estimated $Q$-function
\[
\bb{\hat\pi}_t \in \arg\max_{\bb a_t \in \bb A_t} (\bb\Phi_{t}^\T \bb{\hat\beta}_{t}).
\]
Motivated by the concept of reluctant modeling in the literature, our approach permits some components of $\bb{\hat\theta}$ to be active (nonzero), such that a shift in the corresponding coefficients occurs only when the signal is essential for improving predictive performance, inspired by the principle of reluctant modeling  \citep{yu2019reluctant, tay2020reluctant}. Figure \ref{fig:diagram} below presents the conceptual diagram of the proposed RTL method.

Note that $\lambda_n$ in \eqref{obj} is a tuning parameter that controls the model complexity of $\bb \Phi_t$, and $\bb w = (w_1, \ldots, w_p)$ is a vector of weights used to adjust the level of penalization applied to the individual variables. 
In practice, we propose to use adaptive Lasso \citep{alasso} with perturbed elastic net estimates for weight construction, following \citet{zou2009adaptive}. When specific variables must be retained in the model due to clinical importance or domain knowledge, partial regularization through orthogonality can also be applied, as demonstrated in \citet{oh2020building, oh2022generalization}.

\begin{figure*}[!ht] 
\centering
\includegraphics[scale=.3]{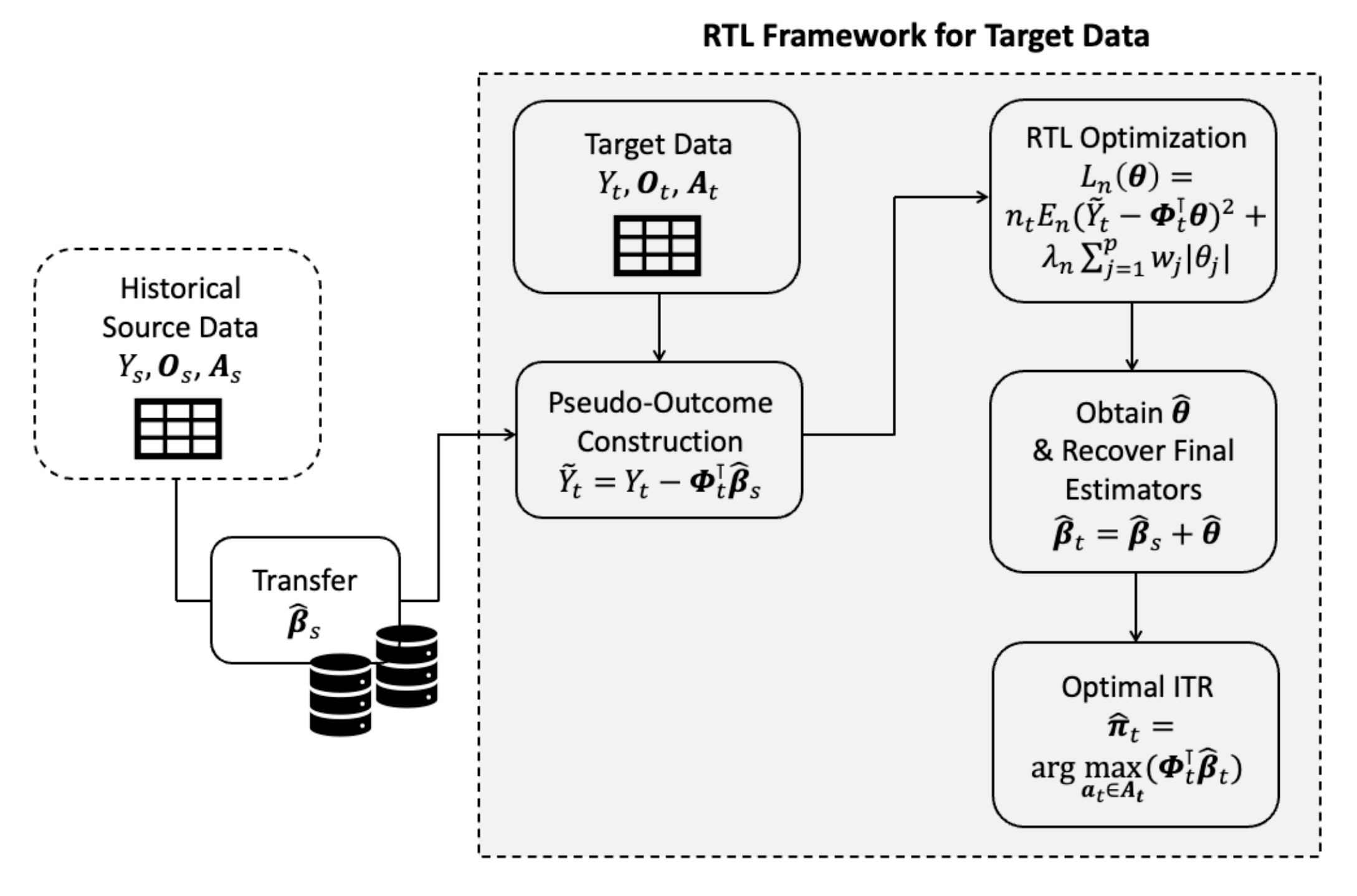}
\caption{A conceptual diagram of the proposed RTL method.}
\label{fig:diagram}
\end{figure*}

Define $\bb \beta_t^* = \arg \min_{\bb \beta_t \in \mathbb{R}^p} E (Y_{t} - \bb \Phi_{t}^\T \bb\beta_t )^2$, and assume that $\bb \beta_t^*$ is unique. Let $\bb \beta_s^*$ be similarly defined. 
In the theorem below, we provide a regret bound for the difference in value of the optimal ITR and that of the estimated ITR.
\noindent Throughout, we assume the following regularity conditions:

\begin{enumerate}[label=(A\arabic*)]
	
    \item \label{a1}
    $\varepsilon_t \overset{\Delta}{=} Y_t - \bb{\Phi}_t^\T\bb\beta_t^*$ has mean zero 
    and finite variance $\sigma^2$.

    \item \label{a2}
    $\bb \Phi_t$ is uniformly bounded.

    \item \label{a3}
    There exist positive constants $b$ and $B$ such that 
    $ b \leq \lambda_{\text{min}}(\bb\Sigma) \leq \lambda_{\text{max}}(\bb\Sigma) \leq B$, 
    where $\bb\Sigma = E(\bb\Phi_t \bb\Phi_t^\T)$, and 
    $\lambda_{\text{min}}(\bb\Sigma)$ and $\lambda_{\text{max}}(\bb\Sigma)$ are the smallest 
    and the largest eigenvalues of $\bb\Sigma$, respectively.

    \item \label{a4}
    $\left \| \hat{\bb\Sigma} - \bb\Sigma \right \|_F \rightarrow_p 0$, where 
    $\left \| \cdot \right \|_F$ stands for the Frobenius norm.

    \item \label{a5}
    $\lambda_n = o(\sqrt{n_t})$.
    
\end{enumerate}

\noindent \textbf{Remark.} Assumption \ref{a1} rules out bias and heavy-tailed disturbances, while Assumption \ref{a2} prevents instability caused by extreme covariates. Assumption \ref{a3} requires that the matrix $\bb\Sigma$ exhibits reasonably good behavior, and Assumption \ref{a4} is needed in the random design case. In addition, note that convergence in the Frobenius norm implies convergence in the operator norm, which further guarantees the consistency of the sample eigenvalues. When $p^2 / n_t \rightarrow 0$, this condition is usually satisfied (see, for example, the proof of Lemma 3.1 in \citet{ledoit2004well}). Lastly, Assumption \ref{a5} is required to establish the rate of convergence for the value of the estimated ITR. \\

\begin{theorem}
Suppose the conditions \ref{a1}--\ref{a5} hold, and $p(\bb a_t|\bb o_t) \ge S^{-1}$ for a positive constant $S$ for all $(o_t,a_t)$ pairs. Assume that there exists some constants $C > 0$ and $\eta \ge 0$ such that
\begin{align}
\bb{P} \left( \max_{\bb a_t \in \mathcal{A}_t} Q_t^o (\bb O_t, \bb a_t)  - \max_{\bb a_t \in \mathcal{A}_t / \arg \max_{\bb a_t \in \mathcal{A}_t} Q_t^o (\bb O_t, \bb a_t)} 
Q_t^o (\bb O_t, \bb a_t) \le \epsilon \right) \le C \epsilon^{\eta}
\label{eqn:noise}
\end{align}
for all positive $\epsilon$. Then 
\begin{align}
V(\bb{\pi}_t^o) - V(\bb{\hat{\pi}}_t) \leq \tilde{C}' \Big [ E [\bb\Phi_t^\T \bb{\beta}_t^* - Q_t^o]^2 + O_P \Big( \frac{p}{\min(n_t, n_s)} \Big) \Big ]^{(1+\eta)/(2+\eta)},
\label{ub}    
\end{align}
where $\tilde{C}' = \left( 2^{3+4\eta}\, S^{1+\eta}\, C \right)^{1/(2+\eta)}.$
\label{thm1}
\end{theorem}

\noindent \textbf{Remarks.}
\begin{itemize}
    \item[1.] Condition \eqref{eqn:noise} is a margin type condition, which measures the difference in mean outcomes between the optimal action(s) and the best suboptimal action(s) at $\bb O_t$. Condition \eqref{eqn:noise} always holds for all $\epsilon > 0$ for $C = 1$ and $\eta = 0$. See \citet{qian2011performance} for discussion of this condition. 
    \item[2.] The first term in the regret bound in \eqref{ub} is called the approximation error, and the second term is the estimation error, which provides the convergence rate. In addition, if $Q_t^o = \bb\Phi_t^\T \bb{\beta}_t^*$, the inequality \eqref{ub} implies that $V(\bb{\pi}_t^o) - V(\bb{\hat{\pi}}_t) \leq O_P [ ( p/\min(n_t, n_s) ) ]^{(1+\eta)/(2+\eta)}$.
\end{itemize}

\section{Simulation}
\label{sec3}

We conducted a comprehensive simulation study to evaluate the performance of the proposed method (Reluctant Transfer Learning, RTL), comparing it against several benchmarks under varying data-generating mechanisms. The primary goal was to assess how well different approaches estimate optimal treatment rules in a target population, accounting for potential heterogeneity between a source domain and a target domain.

\subsection{Binary treatment setting}
\label{subsec:binary}

\subsubsection{Simulation setup}
\label{subsubsec:setup}

Each simulation consisted of a source dataset with $n_s \in \{50, 150\}$ observations and a target dataset with $n_t = 30$ observations. An independent test dataset of $n_{\text{test}} = 2,000$ observations was used to evaluate performance. Covariates $\bb O_s, \bb O_t \in \mathbb{R}^q$ were generated from a multivariate normal distribution with mean zero and autoregressive covariance $\Sigma_{\text{AR}}(\rho)$, where $q = 20$ and $\rho = 0.3$. Treatment assignment $\bb A_s, \bb A_t \in \{0, 1\}$ was independently drawn from a Bernoulli(0.5) distribution.

Various methods were implemented and compared to evaluate their ability to capture domain-specific effects. In the RTL approach, an adaptive Lasso model was fitted to the source data, and the resulting regression coefficients were used to construct a pseudo-outcome. The shift parameters were then estimated to recover the regression coefficients specific to the target data. We considered $\bb \Phi_t = (1, \bb A_t, \bb O_t, \bb A_t {\color{black}{\otimes}} \bb O_t) \in \mathbb{R}^{2q+2}$ for the working model. For comparison, we considered a model that does not transfer any knowledge and develops a model using the target source only (TargOnly) and the integrative transfer learning (ITL) method proposed by \citet{wu2023transfer}. The ITL framework aims to estimate optimal individualized treatment rules by combining covariate balancing weights with outcome modeling. As an additional benchmark, we included the TransLasso method proposed by \citet{li2022transfer}, which is a transfer learning algorithm that first estimates parameters using \textcolor{black}{both auxiliary (source) and primary (target) data} with Lasso, and then corrects the bias using the primary data to improve estimation accuracy.

Two cases of $\bb \beta_s$ were considered, namely (i) weak dense and (ii) sparse signal, both adapted from \citet{oh2020building}, with an effect size of 0.5. In the weak dense case, the source coefficients corresponding to the main effects of covariates were specified as $\{1.25_{q/2}, 0_{q/2}\}$. In the sparse signal case, they were defined as $\{\text{seq}(.1q + .5, 1.5)_{.1q}, 0_{.9q}\}$. In both cases, the outcome models for the source and target populations were defined using structured coefficients, denoted by $\bb \beta_s$ for the source and $\bb \beta_t = \bb \beta_s + \bb \theta$ for the target. The source coefficients captured the main effects and interactions between treatment and covariates. Target-specific shifts were then introduced via distinct $\bb \theta$ vectors in three configurations, each reflecting different levels of heterogeneity and complexity. The following three scenarios for the shift vector $\bb \theta$ were considered:

\begin{itemize}
    \item \textbf{Scenario I}:  A moderate shift in a few main and interaction effects. Specifically, the components of $\bb\theta$ corresponding to the main and its interaction terms for the first two covariates are set to 2.5, while all other components are set to 0.
    
    \item \textbf{Scenario II}: A broader interaction effect shift with strong early magnitude. Specifically, the components of $\bb\theta$ corresponding to the interaction terms are set to $\{\text{seq}(4.5, 1.5)_{.3q}, 0_{.7q}\}$, and all other components are set to 0.

    \item \textbf{Scenario III}: A minimal shift in a majority of coefficients, including 70\% for the interaction terms. Specifically, the components of $\bb\theta$ corresponding to the interaction terms were set to $(0.75_{.7q}, 0_{.3q})$, and all other components were set to 0.75.

\end{itemize}

To evaluate model performance, several metrics were computed on each simulated test dataset. The value was used to quantify the average potential outcome achieved under the estimated optimal treatment rule, specifically within the target population. 
Variable selection performance was evaluated mainly for the interaction terms between treatment and observed covariates based on the number of true zero coefficients correctly estimated as zero (C), and the number of true zero coefficients incorrectly estimated as nonzero (IC).
In addition, the root mean squared error (RMSE) was used to assess prediction accuracy by comparing the model’s predicted outcomes against the actual outcomes in the test data. Each method was evaluated over $100$ replications.

\subsubsection{Simulation results}

Table \ref{table:scenariosI-III} summarizes the performance metrics under the weak dense case. Across all scenarios and sample sizes, the RTL method consistently produced value estimates that are closest to the optimal values. For example, RTL achieved values of 6.47, 7.25, and 6.04 in Scenarios I, II, and III, respectively, when $n_s = 50$. As the sample size increased to $n_s = 150$, the corresponding values increased to 6.58, 7.35, and 6.08, which are close to the optimal values of 6.61, 7.42, and 6.41, respectively. When a minimal shift structure was present (Scenario III), the RMSE of the ITL method was slightly smaller than that of RTL. Nonetheless, RTL maintained competitive value estimation while achieving the highest value with the lowest or comparable RMSE. Moreover, RTL demonstrated favorable variable selection performance, showing high true negative rates (C) and low false inclusions (IC) across settings. In contrast, other methods yielded less accurate value estimates and higher RMSEs, particularly under moderate or strong shifts. \textcolor{black}{While the TransLasso method showed poor variable selection performance, it performed relatively well compared with RTL in terms of value, particularly in Scenario III, where only a minimal shift was observed. However, other aspects, such as variable selection performance and RMSE, suggest that our RTL method outperformed TransLasso. For instance, when $n_s = 50$, C was higher (5 vs 4), IC was lower (1 vs 2), and RMSE was lower (3.91 vs 5.18). Similar patterns were observed as the size of the source data increased to $n_s = 150$.}

\begin{table}[!ht]
  \begin{center}
  \caption{Simulation results under the weak dense case. The mean value is reported with the standard deviation in parentheses. The median C, IC, and RMSE are recorded with the mean absolute deviation in parentheses. The best results are highlighted in boldface.}
\label{table:scenariosI-III}
\begin{threeparttable}
  \fontsize{10}{10}\selectfont
  \begin{tabularx}{0.98\textwidth}{ll cccc cccc}
    \toprule
         &  & \multicolumn{4}{c}{$n_s=$ 50} & \multicolumn{4}{c}{$n_s=$ 150} \\
    \cmidrule{3-10} 
    & Method & Value & C & IC & RMSE & Value & C & IC & RMSE \\
    \midrule
    \multicolumn{2}{l}{Scenario I} \\
    & Optimal & 6.61 & & & & 6.61 & & & \\
    & RTL & \textbf{6.47} (0.10) & 10 (0) & 0 (0) & \textbf{1.49} (0.44) & \textbf{6.58} (0.03) & 10 (0) & 0 (0) & \textbf{0.76} (0.24) \\
    & TargOnly & 5.96 (0.47) & 9 (1) & 1 (1) & 4.17 (1.48) & 6.02 (0.42) & 9 (1) & 1 (1) & 3.99 (1.48) \\
    & ITL & 4.05 (0.45) & 0 (0) & 10 (0) & 2.54 (0.01) & 4.52 (0.58) & 0 (0) & 10 (0) & 2.54 (0.02) \\
    & TransLasso & 6.30 (0.16) & 6 (3) & 4 (3) & 3.29 (0.66) & 6.52 (0.06) & 8 (2) & 2 (2) & 2.23 (0.43) \\

    \multicolumn{2}{l}{Scenario II} \\
    & Optimal & 7.42 & & & & 7.42 & & & \\
    & RTL & \textbf{7.25} (0.22) & 9 (1) & 1 (1) & \textbf{1.90} (0.72) & \textbf{7.35} (0.13) & 10 (0) & 0 (0)  & \textbf{1.08} (0.42) \\
    & TargOnly & 6.79 (0.54) & 9 (1) & 1 (1) & 5.22 (2.17) & 6.83 (0.51) & 9 (1) & 1 (1) & 4.92 (1.98)\\
    & ITL & 3.51 (0.91) & 0 (0) & 10 (0) & 3.54 (0.02) & 4.51 (1.17) & 0 (0) & 10 (0) & 3.54 (0.02)\\
    & TransLasso & 7.04 (0.26) & 6 (1) & 4 (1) & 4.21 (1.19) & 7.27 (0.16) & 7 (3) & 3 (3) & 2.76 (0.63) \\
    
    \multicolumn{2}{l}{Scenario III} \\
    & Optimal & 6.41 & & & & 6.41 & & & \\
    & RTL & \textbf{6.04} (0.14) & 5 (1) & 1 (1) & 3.91 (0.61) & 6.08 (0.14) & 5 (1) & 1 (1) & 3.89 (0.71) \\
    & TargOnly & 5.25 (0.56) & 5 (1) & 1 (1) & 6.77 (1.44) & 5.32 (0.46) & 5 (1) & 1 (1) & 6.60 (1.21) \\
    & ITL & 4.07 (0.43) & 0 (0) & 6 (0) & \textbf{3.14} (0.02) & 4.65 (0.72) & 0 (0) & 6 (0) & \textbf{3.14} (0.02)\\
    & TransLasso & 5.95 (0.18) & 4 (1) & 2 (1) & 5.18 (0.68) & \textbf{6.10} (0.12) & 3 (1) & 3 (1) & 4.64 (0.58) \\

    \bottomrule
  \end{tabularx}
  \end{threeparttable}
  \end{center}
\end{table}

The simulation results under the sparse signal case are presented in Table \ref{table:scenariosIV}. The RTL method produced value estimates that were closest to the optimal values, with relatively low RMSE across various scenarios and sample sizes. \textcolor{black}{Unlike the weak dense case, the advantage of the RTL method in terms of value was less evident, as either TargOnly or TransLasso showed comparable value estimates. However, both TargOnly and TransLasso did not consistently produce values close to the optimal ones across different scenarios. Specifically, TargOnly achieved a high value in Scenario I but not in Scenario III, whereas TransLasso achieved a high value in Scenario III but not to the same extent in Scenario II. In particular, a high value in Scenario III by TransLasso is somewhat expected since when a minimal shift is present, leveraging both sites from the beginning could be a viable option, because TransLasso uses both source and target samples before correcting the bias for the target samples later. However, this comes with a caveat, as TransLasso tended to have a high RMSE (e.g., 5.16 in Scenario III under $n_s = 50$). On the contrary, our method consistently demonstrated the highest value with low RMSE across all scenarios.} Furthermore, the ITL method yielded either very low value or high RMSE, or both. The results highlight the ability of RTL to learn optimal treatment rules with strong predictive performance across a range of scenarios and signal structures.

\begin{table}[!ht]
  \begin{center}
  \caption{Simulation results under the sparse signal case. The mean value is reported with the standard deviation in parentheses. The median C, IC, and RMSE are recorded with the mean absolute deviation in parentheses. The best results are highlighted in boldface.}
\label{table:scenariosIV}
\begin{threeparttable}
  \fontsize{10}{10}\selectfont
  \begin{tabularx}{1\textwidth}{ll cccc cccc}
    \toprule
         &  & \multicolumn{4}{c}{$n_s=$ 50} & \multicolumn{4}{c}{$n_s=$ 150} \\
    \cmidrule{3-10} 
    & Method & Value & C & IC & RMSE & Value & C & IC & RMSE \\
    \midrule
    \multicolumn{2}{l}{Scenario I} \\
    & Optimal & 5.97 & & & & 5.97 & & & \\
    & RTL & \textbf{5.95} (0.06) & 18 (0) & 0 (0)  & \textbf{0.60} (0.21) & \textbf{5.95} (0.03) & 18 (0) & 0 (0) & 0.58 (0.25)\\
    & TargOnly & 5.93 (0.04) & 18 (0) & 0 (0) & 0.74 (0.33) & 5.93 (0.03) & 18 (0) & 0 (0) & 0.63 (0.25)\\
    & ITL & 3.32 (0.51) & 0 (0) & 18 (0) & 2.03 (0.02) & 3.74 (0.85) & 0 (0) & 18 (0) & 2.03 (0.02)\\
    & TransLasso & 5.84 (0.12) & 16 (1) & 2 (1) & 2.51 (0.59) & 5.91 (0.07) & 16 (1) & 2 (1) & 2.08 (0.42) \\
    
    \multicolumn{2}{l}{Scenario II} \\
    & Optimal & 6.51 & & & & 6.51 & & & \\
    & RTL & \textbf{6.38} (0.25) & 14 (0) & 0 (0) & \textbf{1.16} (0.67) & \textbf{6.42} (0.17) & 14 (0) & 0 (0) & \textbf{1.00} (0.50)\\
    & TargOnly & 6.31 (0.33) & 14 (0) & 0 (0) & 1.91 (0.78) & 6.35 (0.29) & 14 (0) & 0 (0) & 1.77 (0.63)\\
    & ITL & 2.18 (0.93) & 0 (0) & 14 (0) & 2.91 (0.02) & 2.80 (1.30) & 0 (0) & 14 (0) & 2.91 (0.01)\\
    & TransLasso & 6.16 (0.29) & 10 (1) & 4 (1) & 3.71 (0.93) & 6.34 (0.20) & 12 (1) & 2 (1) & 2.70 (0.61) \\
    
    \multicolumn{2}{l}{Scenario III} \\
    & Optimal & 5.03 & & & & 5.03 & & & \\
    & RTL & \textbf{4.56} (0.16) & 5 (0) & 1 (0) & 3.76 (0.64) & \textbf{4.57} (0.18) & 5 (1) & 1 (1) & 3.82 (0.64)\\
    & TargOnly & 4.26 (0.29) & 5 (1) & 1 (1) & 4.70 (0.74) & 4.30 (0.29) & 5 (1) & 1 (1) & 4.77 (0.71) \\
    & ITL & 3.00 (0.38) & 0 (0) & 6 (0) & \textbf{2.13} (0.01) & 3.21 (0.53) & 0 (0) & 6 (0) & \textbf{2.14} (0.01)\\
    & TransLasso & 4.52 (0.20) & 5 (1) & 1 (1) & 5.16 (0.70) & 4.56 (0.17) & 5 (1) & 1 (1) & 5.03 (0.50) \\
    
    \bottomrule
  \end{tabularx}
  \end{threeparttable}
  \end{center}
\end{table}

{\color{black}

\subsubsection{Impact of disjoint support sets on performance}
\label{subsec:impact1}

We considered a scenario with disjoint support sets in the interaction terms between source and target domains. This configuration represents a case in which source model structures are fundamentally misaligned with the target domain in terms of active interaction terms; that is, baseline covariates that were critical for formulating the optimal ITR in the source model no longer carry any signal in the target model, while some of the covariates that previously carried no signal in the source model become informative in the target model. Thus, we set the last 10\% of the interaction terms as active in the target model; these terms previously carried no signal in either the weak dense or sparse signal cases of the source model. Detailed descriptions of this scenario are provided in the Supplementary Materials (Section S.1).

Intuitively, when the sets of important variables for constructing the optimal ITR are entirely non-overlapping, relying on source knowledge can be counterproductive. Under this scenario, all transfer learning-based methods exhibited a decline in performance to some extent due to the misalignment of support sets (Supplementary Table 1). Among them, ITL was most severely impacted, showing the largest performance drop. While RTL and TransLasso also showed some degradation, RTL demonstrated better performance than TransLasso, particularly in the sparse signal case. For instance, when $n_s = 50$, RTL achieved a value of 5.42, outperforming other transfer learning methods, such as ITL (3.10) and TransLasso (4.86). In this sparse signal setting, only the first few interaction terms are nonzero in the source model, so the misalignment between source and target models is expected to be smaller than in the weak dense setting. Our findings demonstrate that RTL remains relatively more robust against severe degradation under such settings, further distinguishing it as a stable alternative to existing transfer learning approaches for ITR estimation. Nevertheless, transfer learning-based methods should be approached with caution when clinical knowledge or prior studies indicate strong evidence of non-overlapping active interaction terms.

\subsubsection{Impact of source model misspecification}
\label{subsec:impact2}

We conducted additional simulations to further illustrate the impact of a misspecified source model. To examine robustness under model misspecification, we generated outcomes for the source data from a model that additionally involved a second-order treatment-covariate interaction with the first two baseline covariates, i.e., $\bb A_s \otimes O_{s1} O_{s2}$, with a corresponding coefficient of 1.5, while repeating the model development and evaluation as in \ref{subsubsec:setup}. 
Under this setting, the performances of RTL, ITL, and TransLasso decreased (Supplementary Tables 2--3), although in most cases the reductions were modest. In particular, under the weak dense case in Scenario II with $n_s = 150$, the value for ITL decreased from 4.51 (Table 1) to 4.09 (Supplementary Table 2), suggesting a non-negligible impact of source model misspecification. In contrast, the decrease was marginal for RTL (7.35 vs.\ 7.31) and TransLasso (7.27 vs.\ 7.24). ITL appears to be more sensitive to source model misspecification because it assumes that the conditional treatment effect remains stable across the source and target models, an assumption that is substantially violated under Scenario II and may be further exacerbated when the source model is misspecified.

\subsection{Multi-armed setting}

Additional simulations are conducted to evaluate our proposed method in comparison to other methods with more than two treatment options. The data-generating distribution and setup resemble those in Section \ref{subsec:binary}, with a few changes tailored to the multi-armed setting. Specifically, the treatment is assigned by randomly allocating each individual to one of three treatment options $\{0, 1, 2\}$ with an equal probability of $1/3$. Following dummy indicator encoding, this assignment is represented as a two-dimensional vector $\bb{A}_t = (A_{t,1}, A_{t,2})$, where treatment option 0 serves as the control (i.e., reference), while $(1,0)$ and $(0,1)$ represent treatment options 1 and 2, respectively. The treatment $\bb{A}_s$ is similarly defined. We specified a pairwise effect size of 0.25 for the $(1,0)$ and $(0,1)$ groups relative to the control group. Since the ITL method only works with a binary treatment setting, we compared performance of the rest of the methods. 
For all Scenarios I--III, the shift vector patterns originally defined for the binary treatment setting were applied symmetrically to each treatment dimension $(A_{t,1}, A_{t,2})$ to maintain structural consistency. To introduce variation between the optimal interventions, the interaction terms associated with $A_{t,1}$ and $A_{t,2}$ were constructed to have the same magnitude but with the opposite sign. 

Under this simulation setting (Supplementary Tables 4--5), the results remained largely consistent with those in the binary treatment setting, demonstrating that the proposed method consistently achieved the highest value with low RMSE across various scenarios. Furthermore, the TargOnly method yielded both low value and high RMSE. Although TransLasso achieved competitive value when $n_s = 150$ under the weak dense case, its variable selection performance and RMSE were still inferior to those of the proposed method. 
Thus, the proposed RTL method remains overall favorable, with higher C and lower IC and RMSE, reflecting more accurate and reliable estimation.
}

\section{Real Data Application} \label{sec4}

\begin{figure}
\centering

\begin{subfigure}{0.8\linewidth}
    \centering
    \includegraphics[width=\linewidth]{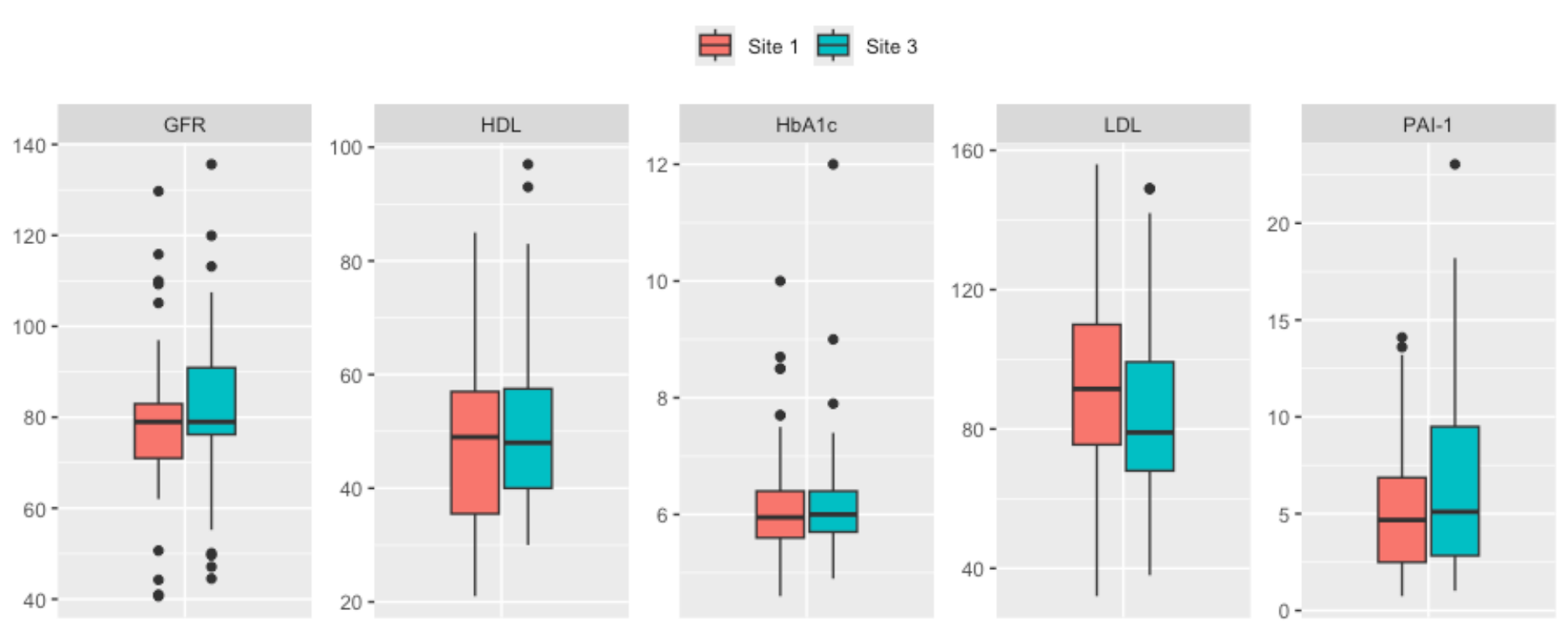}
    \caption{}
    \label{fig:sub1}
\end{subfigure}

\vspace{.5em}

\begin{subfigure}{0.63\linewidth}
    \centering
    \includegraphics[width=\linewidth]{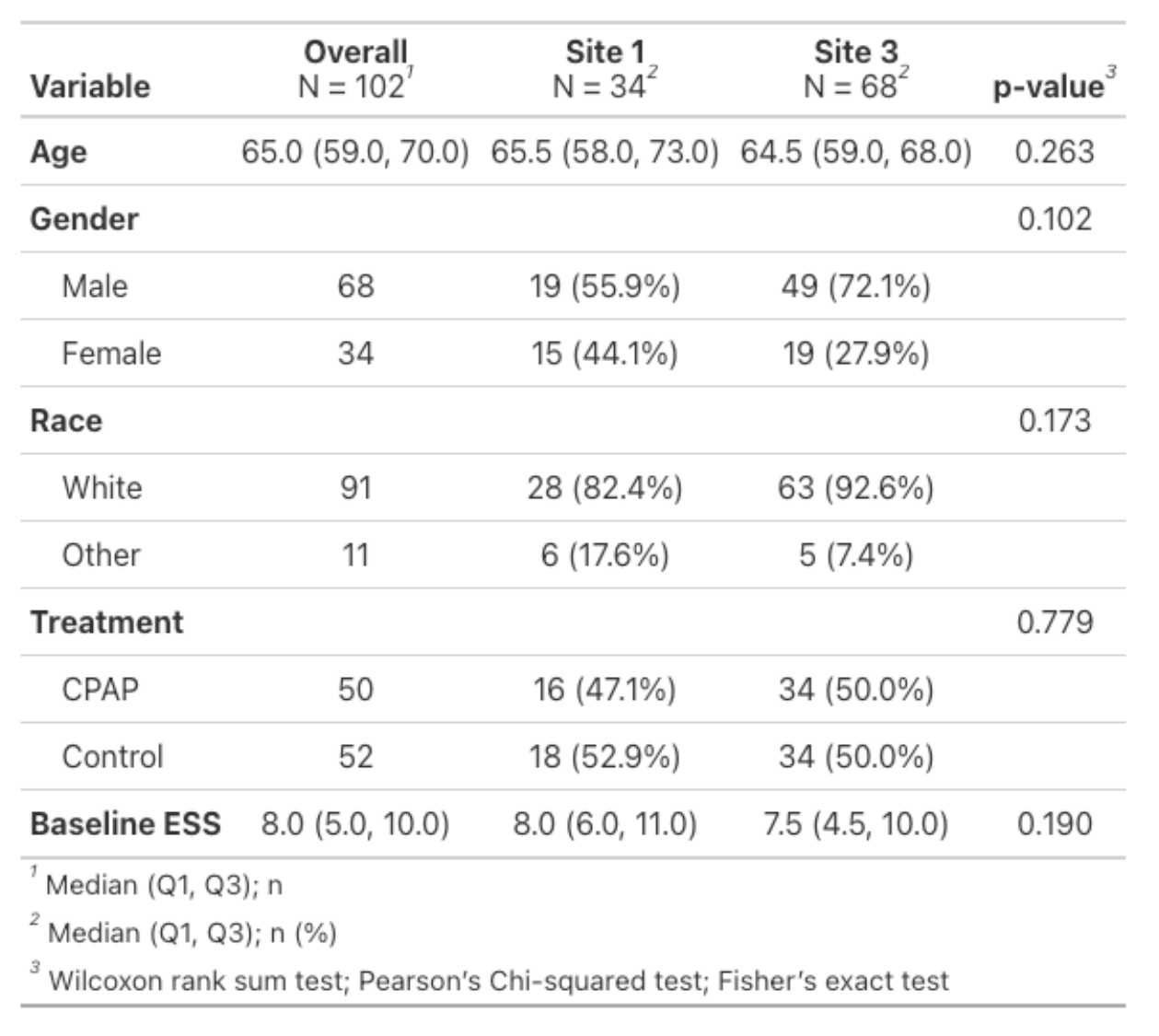}
    \caption{}
    \label{fig:sub2}
\end{subfigure}

\caption{Descriptive analyses of the study cohort: (A) boxplots of several laboratory measurements at baseline across sites; (B) summary statistics in terms of age, gender, race, treatment arm, and baseline ESS (overall and stratified by sites).}
\label{fig:real}
\end{figure}

We applied our proposed method to data from the Best Apnea Interventions for Research (BestAIR) trial, which was designed as a planning study to evaluate key feasibility and study design features within the context of a cardiovascular intervention trial in obstructive sleep apnea (OSA). Following \citet{zhao2017effect}, we analyzed changes in subjective daytime sleepiness, measured by the Epworth Sleepiness Scale (ESS), among participants with cardiovascular comorbidities who were randomized either to the treatment group receiving active continuous positive airway pressure (CPAP) or to the control group receiving conservative medical therapy (CMT) or CMT with sham CPAP. Of the 169 participants, 108 were sucessfully followed for 12 months.  
Our goal was to estimate the optimal regime tailored to each individual in settings where information learned from one site is transferred to improve analyses at another site with limited data. For this purpose, Site 3 was considered the source sample with $n_s = 68$, and Site 1 served as the target sample with $n_t = 34$. Site 2 had extremely limited data with only $6$ participants and was therefore excluded from the analyses. Thus, a total of 102 subjects from Site 1 and Site 3 were included for the analyses.
The primary outcome was the 12-month change in ESS. We considered 25 baseline covariates, including age, gender, race, baseline ESS, and laboratory measurements collected at baseline. Figure \ref{fig:real} presents the clinical and demographic characteristics of the study cohort. 



We applied our proposed RTL method and compared with other existing methods as in Section \ref{subsec:binary}. 
The 3-fold cross-validation was utilized on the target samples \textcolor{black}{(Site 1)} of size $n_t$ to create a one-third hold-out test set for each fold. The remaining two-thirds of $n_t$ samples, along with $n_s$ source samples \textcolor{black}{(Site 3)}, constituted the training set, except for \textcolor{black}{TargOnly where only the remaining two-thirds of $n_t$ samples constituted the training set}. We developed a model on each training set and evaluated the performance metrics on each test set. {\color{black} The value and size of ITR was then averaged across splits to summarize overall model performance, with the mean for value and the median for size of ITR.
The size of ITR is defined as the total number of non-zero coefficients in the treatment-involving component of the model, representing the complexity of the operationalized decision rule. We calculated 95\% confidence intervals for the difference $\Delta$ in estimated values relative to RTL using 1,000 bootstrap resamples, taking the 2.5th and 97.5th percentiles as the interval bounds.} 
The last {\color{black} row} of the tables reports the observed outcome in the data set, which reflects the performance of the treatment allocation used in the study.

Table \ref{table:real} shows that the RTL method achieved the highest estimated value among all methods, {\color{black} suggesting improved treatment recommendation performance. On the other hand, ITL showed inferior performance, yielding the lowest value. This is likely due to its vulnerability to effect heterogeneity, as ITL assumes a stable conditional treatment effect across domains. Both RTL and TargOnly yielded an ITR size of two, consisting of (i) the treatment main effect and (ii) the interaction between treatment and baseline ESS.}
In contrast, the TransLasso method exhibited over-penalization, resulting in {\color{black} an ITR size} of zero, {\color{black} which led to the observed value}. These results suggest that RTL not only outperforms competing methods in terms of value but also maintains minimal complexity, supporting efficient and interpretable ITR estimation. The findings further highlight the practical advantage of selectively transferring model components, allowing adaptation to new data without overfitting or unnecessary model expansion.

\begin{table}[!ht]
  \centering
  \caption{Estimated value and size of ITR for different methods. The best results are highlighted in boldface.}
  \label{table:real}
  \begin{threeparttable}
  \begin{tabular}{lccc}
  \toprule
  Method & Value & $\Delta$ vs. RTL (95\% CI) & Size of ITR \\
  \midrule
  RTL & \textbf{2.40} & --- & 2 \\
  TargOnly & 1.87 & 0.53 (0.13, 0.92) & 2 \\
  ITL & 0.31 & 2.04 (0.13, 4.19) & 25 \\
  TransLasso & 1.14 & 1.28 (0.26, 2.33) & 0 \\
  Observed & 1.14 & 1.28 (0.26, 2.33) & --- \\
  \bottomrule
  \end{tabular}
  \begin{tablenotes}
    \small
    \item $\Delta$ indicates the difference in estimated values between RTL and other methods. Abbreviations: CI, Confidence Interval.
  \end{tablenotes}
  \end{threeparttable}
\end{table}

\section{Discussion} \label{sec5}

In this work, we proposed the RTL framework for the estimation of ITRs to address shifts in treatment-covariate relationships between source and target datasets. Our method selectively transfers essential model components from a source model and allows shifted effects only if they improve predictive performance on the target data. 
By combining model-level transfer with the reluctant modeling principle, the RTL balances adaptability with parsimony and prevents unnecessary complexity. Additionally, a regret bound for the difference between the value of the optimal ITR and that of the estimated ITR was provided in terms of the approximation and estimation errors. Simulation studies showed that our RTL method achieved a higher estimated value than competing approaches, particularly when the sample size of target samples is limited or when a treatment-covariate interaction effect has shifted. In the real data application, 
{\color{black} our RTL method remained robust by adjusting penalization levels across covariates. We hypothesize that real-world data may involve more complex relationships than those represented in the simulation setting, which could have induced excessive shrinkage for methods relying on the standard Lasso penalization, such as TransLasso. In contrast, our proposed RTL method enables differential shrinkage across covariates through a weight vector, $\bb w$, allowing the model to selectively retain important predictors while mitigating over-shrinkage. In addition, unlike methods that require direct access to source data, RTL can be implemented using only the coefficients of a pre-trained source model, further} illustrating its practical utility {\color{black} in privacy-sensitive or data-restricted environments}.


While the RTL approach has shown promise for efficient and interpretable ITR estimation, it has several limitations. It assumes consistent treatment and covariate definitions between source and target datasets and relies on the source model being reasonably well specified. {\color{black} In particular, if new covariates, such as variables not present in the source data, are observed in the target data, an extended model that accounts for these variables may be necessary. One potential approach to accommodate such target-specific predictors is a two-step estimation strategy, where the predictions from the target model, fitted on the original covariate set, are used as an offset in a subsequent regression focused on the new variables. However, this heuristic relies on the assumption that existing and new covariates are uncorrelated to ensure unbiased estimation. 
In addition, performance may be susceptible to the risk of negative transfer. As shown in our simulations in Section \ref{subsec:impact1} (disjoint support sets) and Section \ref{subsec:impact2} (source model misspecification), all transfer learning-based methods experienced performance degradation to some extent. While RTL remained relatively more resilient than ITL and/or TransLasso, the reduction in value underscores the need to rigorously assess transfer suitability.
In practice, a pre-inspection may be helpful, such as using residual diagnostics to identify source model misspecification. Investigating clinical knowledge is also vital to identify potential misalignments in interaction terms, as these directly impact transfer learning performance for constructing optimal ITRs.}
Moreover, {\color{black} while our current framework adopts an additive shift formulation, more flexible transformations of the form $\bb\beta_t = f(\bb\beta_s)$ could capture complex source-target relationships, representing an interesting avenue for future research. Additionally, other promising directions} include incorporating more flexible {\color{black}nonlinear} models {\color{black}as an alternative to the linear specification of the $Q$-function, as well as} exploring different settings, such as online updating strategies and multi-layer data integration. 

\section*{Acknowledgements}
The Best Apnea Interventions in Research (BestAIR) was supported by the National Heart, Lung, and Blood Institute (1U34HL105277) and a grant from ResMed Foundation. Equipment was donated by ResMed and Philips Respironics. The National Sleep Research Resource was supported by the National Heart, Lung, and Blood Institute (R24 HL114473, 75N92019R002). \textcolor{black}{We thank the reviewers, the handling editor, and the editor-in-chief for their careful review and thoughtful feedback.}

\section*{Funding}
The authors received no specific funding for this work.

\section*{Conflicts of Interest}
The authors declare no conflicts of interest.

\section*{Data Availability Statement}
The data that support the findings of this study are available from 
National Sleep Research Resource (NSRR). Restrictions apply to the availability of these data, which were used under license for this study. Data are available from https://sleepdata.org/ with the permission of NSRR. 
{\color{black} The codes for model estimation and simulation are available on Github: https://github.com/oheunj/RTL.}


\printbibliography

\section*{Appendix}

\noindent \textbf{Proof of Theorem \ref{thm1}}. We first derive the excess loss bound of our estimator in terms of $\|\bb{\hat{\beta}}_t - \bb{\beta}_t^*\|_2^2$, and subsequently establish the convergence rate of the value function associated with the estimated ITR. First, observe that
\begin{align*}
\|\bb{\hat{\beta}}_t - \bb{\beta}_t^*\|_2^2  
& = \|(\bb{\hat{\theta}} + \bb{\hat{\beta}}_s) - (\bb{\theta}^* + \bb{\beta}_s^*)\|_2^2 \\
& \le 2\|\bb{\hat{\theta}} - \bb{\theta}^*\|_2^2 
  + 2\|\bb{\hat{\beta}}_s - \bb{\beta}_s^*\|_2^2.
\end{align*}

\noindent Next, it is worth noting that
\begin{align*}
\|\bb{\hat{\theta}} - \bb{\theta}^*\|_2^2
& \le \big\|\bb{\hat{\theta}}(\lambda_n) - \bb{\hat{\theta}}(0) + \bb{\hat{\theta}}(0) - \bb{\theta}^* \big\|_2^2 \\
& = \Big\|
\big(\bb{\hat{\theta}}(\lambda_n) - \bb{\hat{\theta}}(0)\big)
+ \big(\bb{\beta}_s^* - \bb{\beta}_s\big)
+ \bb{\hat\Sigma}^{-1} E_n [\bb{\Phi}_t^{\!\top} \varepsilon_t]
\Big\|_2^2,
\end{align*}
since $\hat{\bb{\theta}}(0)= \arg\min_{\bb{\theta}}\ n_t E_n\!\left[(Y_t-\bb{\Phi}_t^{\!\top}(\hat{\bb{\beta}}_s+\bb{\theta}))^2\right]$, which satiesfies $0 = -2n_t E_n\!\left[\bb{\Phi}_t (Y_t-\bb{\Phi}_t^{\!\top}\left(\hat{\bb{\beta}}_s+\hat{\bb{\theta}}(0))\right)\right]$ and  $\hat{\bb{\theta}}(0) = (\bb{\beta}_s^*-\hat{\bb{\beta}}_s)+\bb{\theta}^*+\bb{\hat\Sigma}^{-1} E_n[\bb{\Phi}_t \varepsilon_t]$. 
In addition, by applying similar arguments to those used in the proof of Lemma 1 of \citet{oh2020building}, we have
\[
\Big\| \bb{\hat{\theta}}(\lambda_n) - \bb{\hat{\theta}}(0) \Big\|_2^2 \le \frac{\lambda_n^2 \sum_j \hat{w}_j^2 }{4 n_t^2 (\lambda_{\min}(\bb{\hat\Sigma}))^2}.
\]
Therefore,
\begin{align*}
\|\bb{\hat{\theta}} - \bb{\theta}^*\|_2^2
&= \Big\|
(\bb{\hat{\theta}}(\lambda)-\bb{\hat{\theta}}(0))
+ \big(\bb{\beta}_s^*-\bb{\hat{\beta}}_s\big)
+ \bb{\hat\Sigma}^{-1} E_n [\bb{\Phi}_t \varepsilon_t]
\Big\|_2^2 \\[2pt]
&\le 2\Big\|
(\bb{\hat{\theta}}(\lambda)-\bb{\hat{\theta}}(0))
+ \bb{\hat\Sigma}^{-1} E_n [\bb{\Phi}_t \varepsilon_t]
\Big\|_2^2
+ 2\|\bb{\hat{\beta}}_s-\bb{\beta}_s^*\|_2^2 \\
&\le 4\|\bb{\hat{\theta}}(\lambda)-\bb{\hat{\theta}}(0)\|_2^2
  + 4\|\bb{\hat\Sigma}^{-1} E_n [\bb{\Phi}_t \varepsilon_t] \|_2^2
  + 2\|\bb{\hat{\beta}}_s-\bb{\beta}_s^*\|_2^2 \\
&\le 4\|\bb{\hat{\theta}}(\lambda)-\bb{\hat{\theta}}(0)\|_2^2
  + 4 \frac{\| E_n [\bb{\Phi}_t \varepsilon_t] \|_2^2 }{ (\lambda_{\min}(\bb{\hat\Sigma}))^2}
  + 2\|\bb{\hat{\beta}}_s-\bb{\beta}_s^*\|_2^2 \\
&\le 4  \frac{ \lambda_n^2 \sum_j \hat{w}_j^2 + \| n_t E_n [\bb{\Phi}_t \varepsilon_t] \|_2^2}{ n_t^2 (\lambda_{\min}(\bb{\hat\Sigma}))^2 }
  + 2\|\bb{\hat{\beta}}_s-\bb{\beta}_s^*\|_2^2  \\
&\le 4  \frac{\lambda_n^2 p + \| n_t E_n [\bb{\Phi}_t \varepsilon_t] \|_2^2}{ n_t^2 (\lambda_{\min}(\bb{\hat\Sigma}))^2 }
  + 2\|\bb{\hat{\beta}}_s-\bb{\beta}_s^*\|_2^2,
\end{align*}
where we set $\hat{w}_j = 1$ in the last inequality. Note that $E \left \| n_t E_n \bb \Phi_t \varepsilon_t \right \|_2^2 = E ( \sum_{i=1}^{n_t} \bb{\Phi}_{ti} \varepsilon_{ti})^2 = n_t \sigma^2 E (\bb\Phi^\T \bb\Phi) = n_t \sigma^2 \Tr(E (\bb\Phi^\T \bb\Phi)) = n_t \sigma^2 \Tr (\bb \Sigma) \leq n_t \sigma^2 p \lambda_{\text{max}} (\bb\Sigma)$. Thus, 

\begin{align*}
\|\bb{\hat{\beta}}_t - \bb{\beta}_t^*\|_2^2 
& \le 2 \left[ 4  \frac{\lambda_n^2 p + \| n_t E_n [\bb{\Phi}_t \varepsilon_t] \|_2^2}{ n_t^2 (\lambda_{\min}(\bb{\hat\Sigma}))^2 }
  + 2\|\bb{\hat{\beta}}_s-\bb{\beta}_s^*\|_2^2  \right] + 2 \|\bb{\hat{\beta}}_s-\bb{\beta}_s^*\|_2^2 \\
& \le 8 \frac{\lambda_n^2 p + \| n_t E_n [\bb{\Phi}_t\varepsilon_t] \|_2^2}{ n_t^2 (\lambda_{\min}(\bb{\hat\Sigma}))^2 }
  + 6\|\bb{\hat{\beta}}_s-\bb{\beta}_s^*\|_2^2 \\
& = O_P \left(\frac{p}{n_t}\right) + O_P\!\left(\frac{p}{n_s}\right) \\
& \le O_P \left(\frac{p}{\min(n_t,n_s)}\right).
\end{align*}
This implies that our RTL estimator is a root-($\min(n_s, n_t)/p$)-consistent estimator. 

Now we show the convergence rate of the value function of the estimated ITR. Observe that 
\begin{align*}
E [ \bb\Phi_t^\T (\bb{\hat{\beta}}_t - \bb{\beta}_t^*)]^2 
& \leq B \|\bb{\hat{\beta}}_t - \bb{\beta}_t^*\|_2^2 \\
& \leq O_P \left(\frac{p}{\min(n_t,n_s)}\right),
\end{align*}
and note that $E [ \bb\Phi_t^\T \bb{\hat{\beta}}_t - Q_t^o)]^2 \leq 2 E [\bb\Phi_t^\T \bb{\beta}_t^* - Q_t^o]^2 + 2 E [ \bb\Phi_t^\T (\bb{\hat{\beta}}_t - \bb{\beta}_t^*)]^2$. 
Thus, using Theorem 3.1 of \citet{qian2011performance}, we obtain 
\begin{align*}
V(\bb{\pi}_t^o) - V(\bb{\hat{\pi}}_t) 
& \leq C' \Big [2 E [\bb\Phi_t^\T \bb{\beta}_t^* - Q_t^o]^2 + 2 E [ \bb\Phi_t^\T (\bb{\hat{\beta}}_t - \bb{\beta}_t^*)]^2 \Big ]^{(1+\eta)/(2+\eta)} \\
& \leq \tilde{C}' \Big [ E [\bb\Phi_t^\T \bb{\beta}_t^* - Q_t^o]^2 + O_P \Big( \frac{p}{\min(n_t, n_s)} \Big) \Big ]^{(1+\eta)/(2+\eta)},
\end{align*}
where $\tilde{C}' = C' \cdot 2^{(1+\eta)/(2+\eta)} = \left( 2^{3+4\eta}\, S^{1+\eta}\, C \right)^{1/(2+\eta)}.$

\end{document}